\documentclass[aps,prb,twocolumn,superscriptaddress,bibnotes]{revtex4-2} 

\usepackage{amsmath}
\usepackage{amssymb}
\usepackage{amsthm}
\usepackage{bbold,bm}
\usepackage[cal=boondox]{mathalfa}
\usepackage{graphicx}
\usepackage{multirow}

\newcommand\vex[1]{\mathbf{#1}}

\def\ket#1{\mathinner{|{#1}\rangle}}

\def\mod{\;\mathrm{mod}\;}
\usepackage{xcolor}

\begin{document} 

\title{Multiple tunable real-space degeneracies in graphene irradiated by twisted light}

\author{Suman Aich}
\affiliation{Department of Physics, Indiana University, Bloomington, Indiana 47405, USA}

\author{Babak Seradjeh}
\email{babaks@iu.edu.}
\affiliation{Department of Physics, Indiana University, Bloomington, Indiana 47405, USA}
\affiliation{Quantum Science and Engineering Center, Indiana University, Bloomington, Indiana 47405, USA}
\affiliation{IU Center for Spacetime Symmetries, Indiana University, Bloomington, Indiana 47405, USA}

\begin{abstract}

We report the theoretical discovery of multiple real-space degenerate Floquet-Bloch states in monolayer graphene coherently driven by twisted circularly-polarized light. Using Floquet theory, we characterize the real-space structure of quasienergies and Floquet modes in terms of the orbital angular momentum and radial vortex profile of light. We obtain the effective real-space Floquet Hamiltonian and show it supports crossings of Floquet modes, especially at high-symmetry $K$ and $\Gamma$ points of graphene, localized at rings around the vortex center. At specific frequencies, the vortex bound states form a multiply degenerate structure in real-space. This structure is purely dynamically generated and controlled by the frequency and tuned by the intensity of twisted light. We discuss the experimental feasibility of observing and employing multiple real-space degeneracies for coherent optoelectronic quantum state engineering.
\end{abstract}

{
\let\clearpage\relax
\maketitle
}

\section{Introduction}
The proposal to generate quantum Hall states dynamically in irradiated graphene~\cite{Oka_2009} opened a new research area~\cite{Oka_2019,Rudner_2020,Torre_2021} to harness the novel nonequilibrium dynamics of many-body systems, including the experimental realization of Floquet-Bloch bands in the solid state~\cite{Wang_2013,McIver_2019} and synthetic systems~\cite{Rechtsman_2013}, theoretical advances in understanding the quantum thermodynamics of driven quantum systems~\cite{Ponte_2015a, Ponte_2015b, Abanin_2016, Khemani_2016, Else_2016a, Geraedts_2016, Bordia_2017, Else_2016b, Yao_2017, Zhang_2017, Khemani_2017},
and Floquet topological phases in quantum materials~\cite{Lindner_2011, Cayssol_2013, Kundu_2014, Usaj_2014, RodriguezVega_2019, Biao_2020, Ghosh_2020, Nag_2021, Zhu_2021, Zhou_2021, Bomantra_2016, Kim_2019, Wang_2014, Bauer_2019, Cadez_2019, Peng_2020, Zhou_2020, Mondal_2023, Wu_2023}. 
The Floquet topological phases in graphene irradiated by circularly polarized light are characterized by a pair of topological invariants associated with quasienergy gaps at the Floquet zone center and boundary. As the frequency and amplitude of the light vary, the invariants change discontinuously at gap closings. Corresponding to these invariants are midgap modes bound to the edges of the sample that traverse the quasienergy gaps. 
Thus, spatial modulations of the light amplitude or frequency could be used to control the quasienergy gap structure and realize novel optically-tuned functionalities~\cite{RodriguezVega_2019,Katan_2013,Kundu_2016}. 

A dramatic display of spatially modulated light is realized by optical beams carrying nonzero orbital angular momentum (OAM), the so-called twisted light, in which the field amplitude has a radial vortex profile and its phase winds around the propagation axis~\cite{Allen_1992,Allen_1999,Andersen_2006,Schmiegelow_2016,Kwon_2019,Kong_2017,Ji_2020,Turpin_2017}. Both twisted linearly-polarized and circularly-polarized light break time-reversal symmetry and realize a Chern electronic insulator in graphene. However, despite the spatial variation, linear polarization does not induce variations of the local Chern marker ~\cite{Bianco_2011,Kim_2022,Bhattacharya_2022,Session_2023,Ahmadabadi_2022}. Since uniform circular polarization induces topological phase transitions by tuning the light frequency as well as its amplitude~\cite{Kundu_2014}, the question arises whether the spatial vortex profile of \emph{twisted} circularly-polarized light can thus create multiple coexisting topological phases.

In this work, we demonstrate that irradiating graphene with twisted circularly-polarized light can indeed realize multiple Floquet topological phases in the same sample. These phases form concentric rings around the light vortex center with a spatial profile determined by the amplitude and frequency of light. At the edges of the rings, the associated quasienergy gap at the Floquet zone center or boundary closes, resulting in Floquet edge modes. Moreover, we find that rings for different gap closings can be tuned to coincide at an infinite set of frequencies, thus creating multiple real-space degeneracies of Floquet bound states. The radii of these degenerate rings can still be tuned by the optical field amplitude, opening a new avenue for optical quantum engineering and control of many-body electronic states. We provide analytical and numerical evidence for these multiple tunable degeneracies and propose a protocol for quantum state control for Floquet bound states at high-symmetry points of graphene's Brillouin zone.

\section{Setup and Model}
We consider graphene subject to coherent circularly-polarized light carrying orbital angular momentum at normal incidence, with vector potential components in the graphene plane coordinates $x+iy = re^{i\phi}$ satisfying $A_x + i A_y = A_0 f(r) e^{i(\Omega t + m \phi)}$. Here $A_0$ is the amplitude, $\Omega$ is the frequency, the integer $m$ is the OAM, and $f(r)$ is the radial vortex profile of the beam. While the precise form of this profile is Laguerre-Gaussian~\cite{Allen_1999}, we will show analytically that our results do not depend on the specific choice of $f(r)$ as long as it vanishes at the vortex core. For our numerical calculations, we choose a simpler Gaussian form, $f(r) = (r/\xi)^{|m|}e^{-(r/\xi)^2}$, which vanishes with a power $m\neq0$ for $r<\xi$, the vortex size of the order of the wavelength of light. We also assume $\xi\gg a_0 = 0.142$~nm, the graphene lattice spacing, which allows us to treat the position $\mathbf{r}$ ($a_0 \ll r \lesssim \xi$) semiclassically as a parameter in our theory just as $\Omega$ and $A_0$.

We calculate the Floquet eigenstates $u_s^{(n)}(\mathbf{k},\mathbf{r},t)$ and the quasienergies $\varepsilon_s^{(n)}(\mathbf{k},\mathbf{r})$ satisfying the Floquet-Schr\"odinger equation $H_F u_s^{(n)} = \varepsilon_s^{(n)} u_s^{(n)}$, where the Floquet-Bloch Hamiltonian in the sublattice basis of graphene is given as ($\hbar = 1$ in our units),
\begin{equation}\label{eq:FBH}
H_F(\mathbf{k},\mathbf{r},t) = \begin{pmatrix} -i\partial_t & -\gamma Z(\mathbf{k},\mathbf{r},t) \\ -\gamma Z^*(\mathbf{k},\mathbf{r},t) & -i\partial_t \end{pmatrix}.
\end{equation}

\begin{figure}[t]
\includegraphics[width=3in]{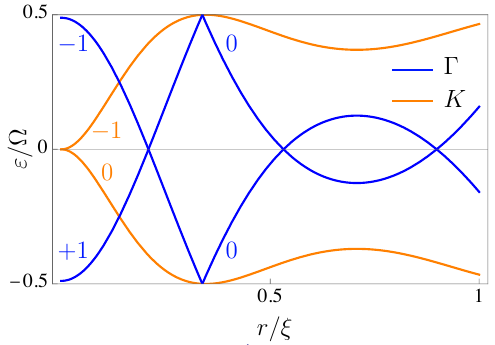}
\caption{Quasienergies as a function of radial position for the $\Gamma$ (blue) and $K$ (orange) points with parameter values: $\Omega/\gamma = 2.0$ and $\alpha = 6$. Labels show the Fourier indices $n_\Gamma$ (blue) and $n_K$ (orange) for each branch.}\label{fig:qen}
\end{figure}

Here, $\mathbf{k}$ is the lattice momentum in graphene's Brillouin zone, $\gamma \ = 2.7$~eV is the nearest-neighbor hopping amplitude of graphene and the tight-binding structure factor $Z(\mathbf{k},\mathbf{r},t) = \sum_{j=1}^3 e^{i[\mathbf{k} + (e/c)\mathbf{A}(\mathbf{r},t)].\mathbf{a}_j}$, where the nearest-neighbor vectors $\mathbf{a}_j = a_0(\cos\theta_j,\sin\theta_j)$, with $\theta_j = (2j-1)\pi/3$. Keeping the lowest order in the Fourier expansion of $Z(t)$ (see Appendix~\ref{app:Floq}), we obtain
\begin{align}
&\varepsilon^{(n)}_\pm(\Gamma) = n\Omega \pm 3\gamma\mathcal{J}_0(\alpha f(r)), \label{eq:eG} \\ 
&u^{(n)}_\pm(\Gamma) = \frac{1}{\sqrt{2}}\begin{pmatrix}1 \\ \mp 1\end{pmatrix} e^{in\Omega t}, \label{eq:uG}
\end{align}
at the $\Gamma$ point, and
\begin{align}
&\varepsilon^{(n)}_\pm(\mathbf{K}) = \bigg(n+\frac{1}{2}\bigg)\Omega \pm \sqrt{\bigg(\frac{\Omega}{2}\bigg)^2 + [3\gamma\mathcal{J}_1(\alpha f(r))]^2}, \label{eq:eK}\\ 
&u^{(n)}_\pm(\mathbf{K}) = \frac{1}{\sqrt{2}}\begin{pmatrix} e^{i(\Omega t+m\phi)}u_{A,B} \\ u_{B,A}\end{pmatrix} e^{i(n\Omega t + \mathbf{K}\cdot\mathbf{r})}, \label{eq:uK}
\end{align}
at the $K$ point, $\mathbf{K} = (0,-1){4\pi}/{3\sqrt{3}a_0}$.
Here, the dimensionless amplitude $\alpha = eA_0a_0/c$,  $\mathcal{J}_n$ is a Bessel function, and $u_{A,B}$ are time-independent functions of position.

Using Eqs.~\eqref{eq:eG} and~\eqref{eq:eK}, we find at the Floquet zone center, the crossings occur between $\pm n_{\Gamma}$ Fourier modes at the $\Gamma$ point and between $n_{K},-n_{K}-1$ modes at the $K$ point, where $n_{\Gamma},n_{K}>0$ satisfy
\begin{align}
|\mathcal{J}_0(\alpha f(r_c))| &= n_{\Gamma} \frac{\Omega}{3\gamma}, \label{eq:FZCG} \\
|\mathcal{J}_1(\alpha f(r_c))| &= \sqrt{n_{K}\left(n_{K}+1\right)} \frac{\Omega}{3\gamma}. \label{eq:FZCK}   
\end{align}
Similarly, at the Floquet zone boundary, we find crossings between $n_\Gamma, 1-n_\Gamma$ Fourier modes at the $\Gamma$ point and $\pm n_K$ at the $K$, with $n_\Gamma,n_K>0$ satisfying
\begin{align}
|\mathcal{J}_0(\alpha f(r_c))| &= \left(n_\Gamma-\frac{1}{2}\right)\frac{\Omega}{3\gamma}, \label{eq:FZBG} \\
|\mathcal{J}_1(\alpha f(r_c))| &= \sqrt{n^2_K-\frac{1}{4}}\frac{\Omega}{3\gamma}. \label{eq:FZBK}
\end{align}

Due to the structure of $Z(t)$ at these high-symmetry points~\cite{Kundu_2014} (see also Appendix~\ref{app:Floq}), avoided crossings are found when the Fourier modes are separated by a multiple of 3. Thus, degenerate states at the $\Gamma$ and $K$ points are found at the Floquet zone center when $n_\Gamma \neq 0 \mod 3$ (except if $n_\Gamma=0$) and $n_K \neq 1 \mod 3$, respectively. Conversely, at the Floquet zone boundary, degenerate states are found when $n_\Gamma \neq 2 \mod 3$ and $n_K \neq 0 \mod 3$.

\begin{table}[t]
    \begin{tabular}{c|cc}
    \hline\hline
     Multiple & $\Gamma$ at FZC & $\Gamma$ at FZB \\[-1.5mm]
     degeneracies& ($n_\Gamma\neq 0 \mod 3$) & ($n_\Gamma\neq 2 \mod 3$) \\
     \hline \rule{0pt}{1.25\normalbaselineskip}%
     \raisebox{1.5mm}{$K$ at FZC} & $ \mathcal{J}_{K/\Gamma} = \frac{\sqrt{n_K(n_K+1)}}{n_\Gamma}$ & $\mathcal{J}_{K/\Gamma} = \frac{\sqrt{n_K(n_K+1)}}{n_\Gamma-1/2}$ \\[-3.5mm]
     ($n_K\neq 1 \mod 3$) & & \\[0.5mm]
     \raisebox{1.5mm}{$K$ at FZB} & $\mathcal{J}_{K/\Gamma} = \frac{\sqrt{n_K^2 - 1/4}}{n_\Gamma}$ & $\mathcal{J}_{K/\Gamma} = \frac{\sqrt{n_K^2-1/4}}{n_\Gamma-1/2}$ \\[-3.5mm]
     ($n_K\neq 0 \mod 3$) & & \\
     \hline\hline
    \end{tabular}
    \caption{The conditions for various multiple degeneracies of $\Gamma$- and $K$-points, either of which can occur at Floquet zone center (FZC) or Floquet zone boundary (FZB). Here, $\mathcal{J}_{K/\Gamma} = |\mathcal{J}_1(\alpha f(r_c)) / \mathcal{J}_0(\alpha f(r_c))|$.}
    \label{demo-table}
\end{table}

In Fig.~\ref{fig:qen}, we plot the quasienergies 
at $\Gamma$ and $K$ points, Eqs.~\eqref{eq:eG} and~\eqref{eq:eK}, in the first Floquet zone. As in Ref.~\onlinecite{Kundu_2014}, the quasienergy spectrum contains various band crossings and avoided crossings between different Fourier modes leading to multiple quasienergy degeneracies; however, these degeneracies can now occur for a fixed value of frequency and light amplitude in the same irradiated graphene setup at different positions on the lattice. As seen for the parameters chosen in Fig.~\ref{fig:qen}, the degeneracies at different lattice momenta can also occur at the same position. We will explore this interesting possibility further below.

We note that multiple band degeneracies are due to the radial vortex profile $f(r)$ that ensures spatial variations of the light amplitude in the sub-wavelength scale. The case without OAM, $m=0$, is similar to the uniform circular polarization case~\cite{Kundu_2014}, as is evident from the form of the vector potential. One could ask whether similar effects could also arise for $m=0$ via spatial variations of light amplitude. However, such variations are limited by the diffraction limit either to occur over many wavelengths (in the far field) or to be small in the sub-wavelength scale (in the near field). Thus, the vortex profile concomitant with $m\neq0$ is necessary for the effects discussed in this work.

\section{Effective Floquet Hamiltonian}
To describe the real-space electronic structure near quasienergy band crossings, we project the Floquet-Bloch Hamiltonian onto the subspace of the two degenerate modes around $\Gamma$ and $K$ points to obtain an effective Hamiltonian which has the general form,
\begin{equation}
\tilde{H}_F(\mathbf{k},\mathbf{r}) = \mathbf{d}(\mathbf{k},\mathbf{r})\cdot\bm\sigma + \mu(k, r)\sigma_z,
\end{equation}
where the Pauli matrices $\bm\sigma$ act on the pseudospin subspace of the states $u_\pm$ at the crossing,
and $\mu$ is a dynamically generated mass term that depends only on the radial coordinate $k$ and $r$. Here $\mathbf{k}$ is measured from either $\Gamma$ or $K$ point (not spanning the entire Brillouin zone). Exactly at the $\Gamma$ and $K$ points, $\mathbf{d} = 0$ and $\mu(0,r)$ is nothing but $\varepsilon^{(n)}_\pm(\Gamma)$ or $\varepsilon^{(n)}_\pm(K)$, respectively. Also, as seen from Fig.~\ref{fig:qen}, $\varepsilon^{(n)}_\pm$ and hence $\mu(r)$ vanish and changes sign at a critical radius $r_c$, which results in gap closing either at the Floquet zone center, $n\Omega$, or at the Floquet zone boundary, $(2n+1)\Omega/2$.
Detailed calculations of the effective Hamiltonian are presented in Appendix~\ref{app:Htilde}, where we confirm that, as with untwisted light~\cite{Kundu_2014}, except for the quadratic $\Gamma$-point crossing at the Floquet zone boundary all other crossings are linear.

\begin{figure}[t]
\includegraphics[width=3.5in]{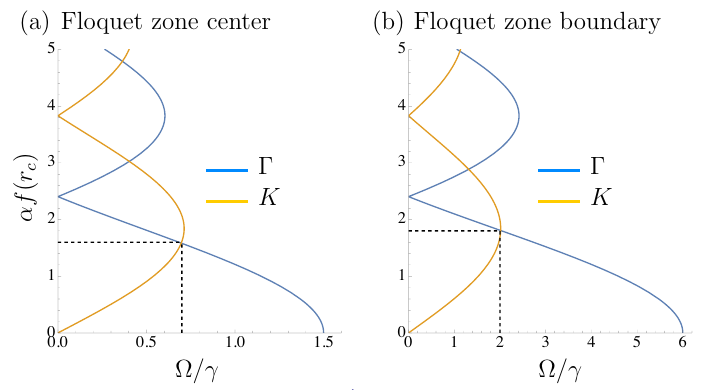}
\caption{Crossings and multiple degeneracies at $\Gamma$ (blue) and $K$ (orange) points at (a) Floquet zone center with $n_\Gamma=n_K=2$, and (b) Floquet zone boundary with $n_\Gamma=n_K=1$. The multiple degeneracies at (a) $\alpha f(r_c) = 1.6$, $\Omega/\gamma \approx 0.7$ and (b) $\alpha f(r_c) = 1.8$, $\Omega/\gamma \approx 2$ are indicated by dashed lines.}\label{fig:MF}
\end{figure}

\begin{figure}[t]
\includegraphics[width=3.2in]{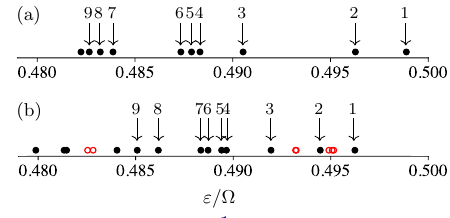}
\caption{Quasienergies near Floquet zone boundary for (a) $\alpha = 3$, $m=1$, $\xi=100a_0$, $\Omega/\gamma = 5.8$, and (b) $\alpha = 22.3$, $m=1$, $\xi=100a_0$~nm, and $\Omega/\gamma = 1$. Analytically, we expect a single $\Gamma$-point crossings for (a) at $r_c = 1.76$~nm and a  $K$-point crossing for (b) at $r_c=2.04$~nm. The black dots and red circles represent bulk and edge bound states respectively. The numbered bulk edge states are presented in Fig.~\ref{fig:uGK}.
} \label{fig:qeGK}
\end{figure}

We find the Floquet eigenstates near band crossings by converting $\mathbf{k}$ to the real-space operator $-i\bm\nabla_{\mathbf{r}}$ and solving the resulting differential equations. The details of our solution are presented in Appendix~\ref{app:wbound}. Here, we note that the effective Hamiltonian commutes with a pseudo-OAM operator $\hat{l}_z$ that combines the orbital and pseudospin angular momentum. Thus, we label the eigenstates of $\tilde{H}_F$ with eigenvalue $l$ of the pseudo-OAM operator, $\chi_{l}(\mathbf{r})^T = (e^{il_+\phi}w_+(r)\;,\;e^{il_-\phi}w_-(r))$,
where again $\pm$ refer to the pseudospin subspace, and $l_\pm$ are combinations of pseudo-OAM, $l$, and pseudospin eigenvalues. The complete Floquet eigenstate in the sublattice basis of graphene takes the form
\begin{equation} \label{eq:psil}
\psi_{l}(\mathbf{r},t) = e^{il_+\phi} w_+(r) u_+(\mathbf{r},t)
+ e^{il_-\phi} w_-(r) u_-(\mathbf{r},t),
\end{equation}
where $u_\pm$ are given by Eqs.~\eqref{eq:uG} and~\eqref{eq:uK} at the corresponding crossing.

We are particularly interested in bound-state solutions for which $w_\pm$ are localized around $r_c$, where the mass term vanishes. These bound states form a ring of radius $r_c$ separating two gapped regions in the graphene sheet. We present analytical properties of these bound states in Appendix~\ref{app:wbound}. 

\begin{figure*}
\includegraphics[width=7.2in]{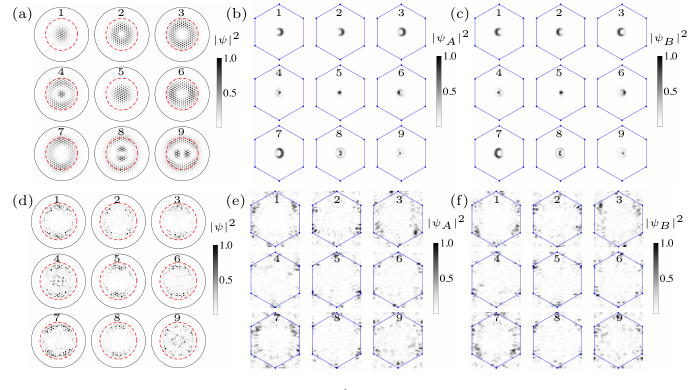}
\caption{Probability distributions in real space $|\psi(\mathbf{r})|^2$ (a and d) and momentum space $|\psi_{A,B}(\mathbf{k})|^2$ (b, c and e, f) for states numbered 1-9 in Fig.~\ref{fig:qeGK}(a) (top row) and Fig.~\ref{fig:qeGK}(b) (bottom row). The red dashed circles show values (a) $r_c = 1.76$~nm and (d) $r_c = 2.04$~nm calculated analytically for $\Gamma$- and $K$-point crossings at the Floquet zone boundary, respectively. The black circles in (a) and (d) denote the edge of the graphene sample. The blue hexagons in (b-c) and (e-f) mark the graphene Brillouin zone.} \label{fig:uGK}
\end{figure*}

\section{Multiple real-space degeneracies}
Remarkably, one can tune the ring of bound states for different crossings to coincide at the same $r_c$. This  realizes multiple real-space degeneracies between various crossings. Using Eqs.~(\ref{eq:FZCG}-\ref{eq:FZBK}), we summarize the general conditions for multiple degeneracies at Floquet zone center and boundary in Table~\ref{demo-table}. Interestingly, since it is the value of $\alpha f(r_c)$ that is fixed by these conditions, the position of the multiple degenerate rings, $r_c$, can itself be tuned by the amplitude $\alpha$.

For example, at the highest frequency, the critical radius for multiple degeneracies of the $\Gamma$- and $K$-point crossings are found at the Floquet zone center with $n_\Gamma=n_K=2$ for $\alpha f(r_c) \approx 1.6$ and $\Omega/\gamma \approx 0.7$, and at the Floquet zone boundary with $n_\Gamma=n_K=1$ for $\alpha f(r_c) \approx 1.8$ and $\Omega/\gamma \approx 2$. For these values of $n_\Gamma$ and $n_K$, we plot $\alpha f(r_c)$ as a function of frequency in Fig.~\ref{fig:MF}. Multiple degeneracies are obtained at the intersection of these plots, giving rise to an infinite set of degenerate frequencies. There is in fact an infinite family of such sets, depending on the degenerate Fourier indices $n_\Gamma$ and $n_K$. This offers a large space of parameters to obtain and tune multiple degenerate states.

Alternatively, real-space degeneracies may also be obtained for one crossing at the Floquet zone center and the other at the Floquet zone boundary. When the former is a $\Gamma$-point crossing we find $n_\Gamma=n_K=1$ at the highest frequency $\Omega/\gamma \approx 1.8$ and $\alpha f(r_c) \approx 1.3$. When the former is a $K$-point crossing we find $n_\Gamma=n_K-1=1$ at the highest frequency $\Omega/\gamma \approx 0.7$ and $\alpha f(r_c) \approx 2.2$. 

We note that these multiple real-space degeneracies are distinguished from the higher-spin momentum-space multifold fermions protected by crystalline symmetries~\cite{Flicker_2018, Schroter_2019, Sanchez_Martinez_2019, Xu_2020} by the fact that they occur at different lattice momenta and possibly even different quasienergies, but are localized in the same region of space.

\section{Numerical Results}
We numerically diagonalize the Floquet time-evolution operator using the full tight-binding Hamiltonian in real space and obtain the Floquet spectrum. We take a graphene sample with a disk geometry of radius $3$~nm $\approx 21 a_0$. The twisted light is implemented as $f(r) = (r/\xi)^{|m|}e^{-(r/\xi)^2}$ with $r$ measured from the center of the disk. We take a large value of $\xi=100a_0$ to explore the Floquet states within the light vortex core. The values of other parameters are chosen so that, based on our analytical expressions, we expect to obtain a band crossing at either the $\Gamma$ point or $K$ point at the Floquet zone boundary within the disk. Note that these values are not representative for experiments, rather they are chosen to allow efficient numerical simulations for relatively small system sizes. The full code generating our results is available online~\cite{aich_2024_10688606}.

Our results are shown in Figs.~\ref{fig:qeGK} and~\ref{fig:uGK}. We find that the states near the Floquet zone boundary shown in Fig.~\ref{fig:qeGK} are indeed localized either within the disk (shown by black dots) or near its edges (red circles). This is so because the two regions separated at $r_c$ are in different Floquet topological phases.

\begin{figure*}
\includegraphics[width=7.2in]{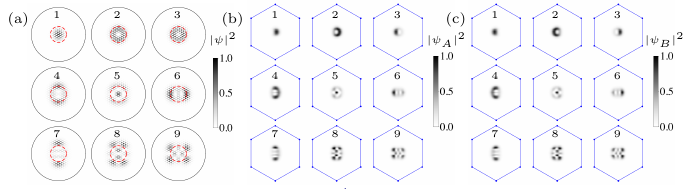}
\caption{Probability distributions in real space $|\psi(\mathbf{r})|^2$ (a) and momentum space $|\psi_{A,B}(\mathbf{k})|^2$ (b, c) for the $\Gamma$-point crossing. Here $\alpha=6$ while other parameters are the same as in Fig.~\ref{fig:qeGK}(a), so that now $r_c=0.88$~nm shown by the red dashed circle in (a).} \label{fig:uG}
\end{figure*}

We present the wavefunctions of several localized states within the sample in Fig.~\ref{fig:uGK}. The real-space probability distributions $|\psi(\vex r)|^2$ shown in Fig.~\ref{fig:uGK}(a) and Fig.~\ref{fig:uGK}(d) clearly show localization near the expected values of $r_c = 1.76$~nm and $r_c=2.04$~nm (shown as red dashed circles) for the $\Gamma$- and $K$-point crossings, respectively. They also show an intricate orbital structure within the sample due to pseudo-OAM eigenvalues, for which we present explicit expressions in Appendix~\ref{app:Htilde}, that results in the variety of shapes of the wavefunctions.

In order to test our analytical results, we find the dominant momentum-space contribution to the bound states by performing a numerical Fourier transform of the wavefunctions separately for each sublattice to obtain $\psi_{A/B}(\vex{k}) = \sum_{\vex r\in A/B} e^{-i \vex k \cdot \vex r} \psi(\vex r)$. The probability distributions $|\psi_A(\vex k)|^2$ and $|\psi_B(\vex k)|^2$ are shown in Fig.~\ref{fig:uGK}(b)-(c) and Fig.~\ref{fig:uGK}(e)-(f). They show clear localization around $\Gamma$ and $K$ points, respectively. Thus, our numerics confirm our analytical findings and provide additional information about the orbital structure of the bound states.

\section{Discussion}
The topological nature of Floquet states we have uncovered here can be seen via Floquet topological invariants~\cite{Kundu_2014}, $C_0(r)$ and $C_\pi(r)$, calculated in the semiclassical approximation in which $\vex r$ is treated as a parameter in the Floquet-Bloch Hamiltonian~\eqref{eq:FBH}. 
For example, for the parameter values as in Figs.~\ref{fig:qeGK}(a) and~\ref{fig:uGK}(a), we have $\alpha f(r)<0.366$ for $r<r_c = 1.76$~nm, which corresponds~\cite{Kundu_2014} to a change from $C_\pi(r<r_c)=-2$ to $C_\pi(r>r_c)=0$ while $C_0(r) = 1$ throughout the sample. The change $\Delta C_\pi(r_c) = 2$ happens through a quadratic band touching at the Floquet zone boundary at the $\Gamma$ point~\cite{Kundu_2014}, in agreement with the effective Hamiltonian derived in Appendix~\ref{app:Htilde}.

Moreover, by bulk-boundary correspondence, a change of the Floquet topological invariants $C_0$ and $C_\pi$ at $r_c$ corresponds to topologically protected bound states localized around $r=r_c$ at, respectively, the Floquet zone center and boundary. This explains the existence of bound states shown in Fig.~\ref{fig:uGK}(a).

Since it is the value $\alpha f(r_c)$ that is fixed for a given frequency, the value of $r_c$ can be changed by tuning $\alpha$. In Fig.~\ref{fig:uG}, we show our numerical results for $\alpha=6$ while keeping the other parameters the same as in Fig.~\ref{fig:uGK}(a), for which $r_c=0.88$~nm. As expected, the bound state wavefunctions are now localized around the predicted value of $r_c$. This proves both the tunability and the topological nature of multiple phases and bound states in the sample due to topologically protected quasienergy degeneracies.

The main property of the twisted light we have focused on so far is the vortex profile $f(r)$, which creates a spatially variable amplitude $\alpha f(r)$ and, therefore, the possibility of concurrent realizations of multiple Floquet topological phases in the same sample. The OAM and angular winding of light also appear in the wavefunctions of the Floquet states, Eqs.~\eqref{eq:uG} and~\eqref{eq:uK}, and the pseudo-OAM eigenvalues in Eq.~\eqref{eq:psil}. 
While the vortex profile is a consequence of $m\neq0$, it is natural to ask if this angular dependence manifests itself more directly. We answer this question in the affirmative by investigating a way to control the quantum states of multiple degenerate bound states associated with $\Gamma$- and $K$-point crossings.

In particular, we propose to manipulate quantum superpositions of the $\Gamma$- and $K$-point multiple bound states through a localized real-space scattering potential $V(\mathbf{r})$ that connects $\Gamma$ to $K$. Such a potential must have some variation between the two sublattices with wave vector $\vex K$, which can naturally exist at the edges of the system or may be created artificially. By tuning the light amplitude $\alpha$, we can tune $r_c$ for multiple degenerate states to coincide with the region where $V(\vex r)\neq 0$. 

As we show in Appendix~\ref{app:selection}, due to the differences in the linear vs. quadratic band crossings, for such a potential to have nonzero matrix elements between $\Gamma$- and $K$-point degenerate states, the light OAM must take values $m=\pm 2$ at Floquet zone center and $m=\pm 1$ at Floquet zone boundary. 
Assuming the scattering potential has a primary wave vector $\vex K$, the dominant scattering process connects $\Gamma$ not only to the $K$-point bound states, but also to the $K'$-point bound states at the opposite valley. The multiple degeneracy conditions and OAM values for $\Gamma\leftrightarrow K'$ and $\Gamma\leftrightarrow K$ transitions are the same. However, the matrix elements for the two transitions can be different from each other and may be switched by the helicity of the circular polarization as well as the sign of $m$.

Therefore, employing OAM offers the possibility to engineer and control two-level systems from multiple degeneracies at the Floquet zone center and boundary. In the ideal situation of nonvanishing couplings $b$ and $b'$, respectively, between $\Gamma$ and $K$, and, $\Gamma$ and $K'$, the two levels are formed from the $\Gamma$-point bound state and a superposition of $K$- and $K'$-point bound states weighted by $b$ and $b'$. Quantum operations can proceed by tuning $r_c$ via the light amplitude $\alpha$: the two levels are stationary as long as $r_c$ does not overlap with the region $V(\vex r) \neq 0$, while a rotation between the two levels is achieved by tuning $r_c$ to overlap with the region $V(\vex r) \neq 0$ for a time $t$, with a precession frequency $\sim \sqrt{b^2 + b'^2}$. Some details of the calculations for the quantum state manipulation process described here are presented in Appendix~\ref{app:trans}. 

The parameters in our simulations were chosen for numerical feasibility and illustration purposes.
For electric field amplitude $2\times 10^8$~V/m in the lab, corresponding to intensities $10^{9}$~W/cm$^2$, we obtain $\alpha \approx 10^{-2}(\Omega/\gamma)^{-1}$. So, for $\Omega/\gamma \approx 0.1$, we can achieve $\alpha \approx 0.1$. Since the vortex profile $f(r)\lesssim 1$, we must have $\alpha \gtrsim \alpha f(r_c)$ to realize multiple degeneracies. For $f(r_c) \lesssim 1$ we find $r_c \sim \xi\sim c/\Omega \approx 5\times10^3 a_0$. Thus, micron-sized samples of graphene can accommodate multiple degeneracies near their edges.
One may utilize higher Fourier modes to enlarge the experimental feasibility of parameters for multiple degeneracies. For example, fixing $n_K=2$ and varying $n_\Gamma>2$ (and $\neq 0 \mod 3$) we obtain the highest-frequency multiple degeneracies for $\alpha f(r_c) \approx 2\sqrt{6}n_\Gamma^{-1}$ and $\Omega/\gamma \approx 3n_\Gamma^{-1}$ at the Floquet zone center. Similarly, fixing $n_K=1$ and varying $n_\Gamma>2$ (and $\neq 2 \mod 3$) we obtain the highest-frequency multiple degeneracies for $\alpha f(r_c) \approx 2\sqrt{3}(2n_\Gamma-1)^{-1}$ and $\Omega/\gamma \approx 6(2n_\Gamma-1)^{-1}$ at the Floquet zone boundary. Taking the Fourier mode $n_\Gamma=10$, we have $\alpha f(r_c) \approx 0.5$ at Floquet zone center and $\alpha f(r_c) \approx 0.2$ at Floquet zone boundary with $\Omega/\gamma \approx 0.3$ in the infrared. 

\begin{figure}[t]
\includegraphics[width=3in]{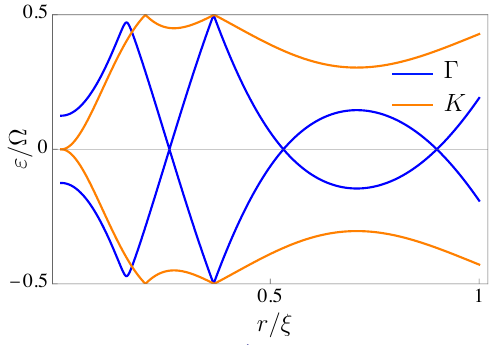}
\caption{Quasienergies as a function of radial position for the $\Gamma$ (blue) and $K$ (orange) points with parameter values: $\Omega/\gamma = 1.6$ and $\alpha = 6$.} \label{fig:S1}
\end{figure}

We have studied the interaction of twisted circularly-polarized light with graphene and shown, both analytically and numerically, that multiple Floquet topological phases can be realized in the same sample. At certain frequencies, multiple real-space degenerate states become possible and may be tuned by light amplitude, thus creating a possible route for \emph{in situ} optical quantum state control of light-matter bound states. Multiple real-space degeneracies occur and can be tuned at subwavelength scale within the light vortex core. By contrast, without OAM, optical field variations can only resolve Floquet topological phases on the scale of many wavelengths. 

Future work on local topological markers of various multiple real-space degeneracies can also characterize their optical and magnetic properties. An interesting possibility to explore is using additional harmonics~\cite{Kundu_2016} of twisted light to control the quantum states of the valleys, which may be operated with graphene leads. 

\begin{acknowledgements}
This work is supported in part by the College of Arts and Sciences and the Institute for Advanced Study at Indiana University. B.S. acknowledges the support of the Aspen Center for Physics, under NSF grant PHY-2210452, where parts of this work were performed.
\end{acknowledgements}
\appendix
\onecolumngrid

\section{Floquet spectra}\label{app:Floq}
Here we calculate the full quasienergies of the Floquet-Bloch Hamiltonian (\ref{eq:FBH}) for both the $\Gamma$ and $K$ points. The time-dependent Hamiltonian is expanded in its Fourier components and organized similar to a time-independent Hamiltonian in the extended Floquet-Hilbert space~\cite{Eckardt_2015,Rodriguez-Vega_2018}.
Using the expression,
\begin{equation}\label{eq:Z}
Z(\mathbf{k},\mathbf{r},t) = \sum_{j=1}^3 e^{i[\mathbf{k} + (e/c)\mathbf{A}(r,\phi,t)] \cdot \mathbf{a}_j} =: \sum_n Z_n(\mathbf{k},\mathbf{r}) e^{-in\Omega t},
\end{equation}
the Fourier components of the tight-binding structure factor 
are given by
\begin{equation}\label{eq:Zn}
Z_n(\mathbf{k},\mathbf{r}) = i^n\mathcal{J}_n(\alpha f(r))\bigg[(-1)^ne^{-ik_xa_0}  + 2e^{ik_xa_0/2}\cos\bigg(\frac{n\pi}{3} + \frac{\sqrt{3}}{2}k_ya_0\bigg)\bigg]e^{-inm\phi},
\end{equation}
where $\mathcal{J}_n$ are Bessel functions, and $\alpha = eA_0a_0/c$ is the dimensionless amplitude. 

Quasienergies obtained here improve the lowest-order expressions in Eqs.~\eqref{eq:eG} and~\eqref{eq:eK} by adding contributions from higher Bessel functions. For numerical purposes we truncate this expansion by keeping a maximum order of Bessel functions set to $n=12$, shown in Fig.~\ref{fig:S1}. As a result, the frequency at which the first multiple real-space degeneracy at Floquet zone boundary, shown in Fig.~\ref{fig:MF}(b), occurs is shifted from $\Omega/\gamma\approx 2$ in the lowest-order approximation to $\Omega/\gamma\approx 1.6$ in the full calculation. This correction is more pronounced for the $K$ point since the next-order Bessel function $\mathcal{J}_2$ contributes more than the next-order Bessel function $\mathcal{J}_3$ at the $\Gamma$ point for a given value of $\alpha f(r) \lesssim 3$.

\section{Effective Hamiltonians at the $\Gamma$ and $K$ points}\label{app:Htilde}
\subsection{Floquet zone center}

Floquet modes with Fourier components $\mp n_\Gamma$ are degenerate at $\Gamma$ and those with Fourier components $n_K,-n_K-1$ can be degenerate at $K$ with $n_\Gamma,n_K>0$ ($n_\Gamma \neq 0 \mod 3$ and $n_K \neq 1 \mod 3$). The degenerate Floquet eigenstates are $ \frac{1}{\sqrt{2}} \begin{pmatrix} 1 \\ \mp 1 \end{pmatrix} e^{\mp in_\Gamma\Omega t}$ for $\Gamma$ and $\frac{1}{\sqrt{2}} \begin{pmatrix} e^{i(\Omega t+m\phi)}u_A \\ u_B \end{pmatrix} e^{i(n_K\Omega t + \mathbf{K\cdot r})} $ and $\frac{1}{\sqrt{2}} \begin{pmatrix} e^{i(\Omega t+m\phi)}u_B \\ u_A \end{pmatrix} e^{-i((n_K+1)\Omega t -\mathbf{K\cdot r})} $ for $K$. Thus, in the degenerate subspace, we find the matrix elements of the effective Hamiltonian at $\Gamma$ are
\begin{align}\label{eq:HG0-11}
(\tilde{H}_F)^{\Gamma(0)}_{11} = - (\tilde{H}_F)^{\Gamma(0)}_{22} &= -n_\Gamma\Omega + 3\gamma\mathcal{J}_0(\alpha f(r))\left(1-\frac{k^2a_0^2}{4}\right), \\
\label{eq:HG0-12}
(\tilde{H}_F)^{\Gamma(0)}_{12} &= i(-1)^{n_\Gamma}\frac{3\gamma}{2}\mathcal{J}_{2n_\Gamma}(\alpha f(r))\left[k_x + i(-1)^{n_\Gamma\mod3}k_y\right]a_0e^{-2in_\Gamma m\phi},
\end{align}
and at $K$ are
\begin{align}\label{eq:HK0-11}
(\tilde{H}_F)^{K(0)}_{11} = - (\tilde{H}_F)^{K(0)}_{22} &= \frac{1}{\eta^2+1}\left[\left((n_K+1)\eta^2 + n_K\right)\Omega + 6\gamma\eta\mathcal{J}_1(\alpha f(r))\left(1 - \frac{k^2a_0^2}{4}\right)\right], \\
 \label{eq:HK0-12}
(\tilde{H}_F)^{K(0)}_{12} &= i(-1)^{(n_K+1)}\frac{3\gamma}{2(\eta^2+1)}\mathcal{J}_{2n_K}(\alpha f(r))(k_x - i\beta k_y)a_0e^{i(2n_K+1)m\phi},
\end{align}
where $\beta = +1$ for $n_K = 0 \mod 3$ and $\beta=-1$ for $n_K = 2 \mod 3$, respectively. We have introduced 
\begin{equation}\label{eq:eta}
\eta(r)= -i \frac{u_A}{u_B} =  \frac{\frac{\Omega}{2} - \sqrt{\left(\frac{\Omega}{2}\right)^2 + \left[3\gamma\mathcal{J}_1(\alpha f(r))\right]^2}}{3\gamma\mathcal{J}_1(\alpha f(r))}.
\end{equation}

Due to the particular angular structure of $\tilde{H}_F^{\Gamma(0)}$ and $\tilde{H}_F^{K(0)}$, we have $[\tilde{H}_F^{\Gamma(0)}, \hat l^{\Gamma(0)}] = 0$ and $[\tilde{H}_F^{K(0)}, \hat l^{K(0)}] = 0$, where $\hat l^{\Gamma(0)}$ and $\hat l^{K(0)}$ are the respective psuedo-OAM operators given by
\begin{align}\label{eq:l0}
   \hat l^{\Gamma(0)} & = -i\partial_\phi + \left(n_\Gamma m - \frac{(-1)^{n_\Gamma\mod3}}{2}\right)\sigma_z, \\
   \hat l^{K(0)} & = -i\partial_\phi - \left[\frac{(2n_K+1)m - \beta}{2}\right]\sigma_z.
\end{align}
Correspondingly, the eigenvalues of the pseudo-OAM operators are
\begin{align}\label{leval0}
&l_\pm^{\Gamma(0)} = l \mp \left(n_\Gamma m - \frac{(-1)^{n_\Gamma\mod3}}{2}\right), \\
&l_\pm^{K(0)} = l \pm \left[\frac{(2n_K+1)m - \beta}{2}\right],
\end{align}
for $\Gamma$ and $K$ respectively. Thus, from the effective Hamiltonians we can see that there is a linear band crossing at both $\Gamma$ and $K$ points. The complete Floquet eigenstate for zero quasi-energy at the $\Gamma$ point can be written as,
\begin{equation}\label{eq:psiG0}
\psi_{l}^{\Gamma(0)}(\mathbf{r},t) = e^{il_+^{\Gamma(0)}\phi}w^{\Gamma(0)}_+(r) u^{(-n_\Gamma)}_+(\Gamma,\mathbf{r},t) \; + \; e^{il_-^{\Gamma(0)}\phi}
w^{\Gamma(0)}_-(r) u^{(n_\Gamma)}_-(\Gamma,\mathbf{r},t).
\end{equation}
Similarly, for the $K$ point
\begin{equation}\label{psiK0}
\psi_{l}^{K(0)}(\mathbf{r},t) = e^{il_+^{K(0)}\phi}w^{K(0)}_+(r) u^{(n_K)}_+(K,\mathbf{r},t) \; +\;
 e^{il_-^{K(0)}\phi}w^{K(0)}_-(r) u^{(-n_K-1)}_-(K,\mathbf{r},t).
\end{equation}

\subsection{Floquet zone boundary}
Now, Floquet modes with Fourier components $1-n_\Gamma,n_\Gamma$ can be degenerate at $\Gamma$ and those with Fourier components $\pm n_K$ at $K$ with $n_\Gamma,n_K>0$ ($n_\Gamma \neq 2 \mod 3$ and $n_K \neq 0 \mod 3$). The degenerate Floquet eigenstates are $ \frac{1}{\sqrt{2}} \begin{pmatrix} 1 \\ - 1 \end{pmatrix} e^{-i(n_\Gamma-1)\Omega t}$ and $\frac{1}{\sqrt{2}} \begin{pmatrix} 1 \\ 1 \end{pmatrix} e^{in_\Gamma\Omega t}$ for $\Gamma$ and $\frac{1}{\sqrt{2}} \begin{pmatrix} e^{i(\Omega t+m\phi)}u_{A,B} \\ u_{B,A} \end{pmatrix} e^{\pm i(n_K\Omega t \pm \mathbf{K.r})} $ for $K$. Thus, in the degenerate subspace, the matrix elements of the effective Hamiltonian at $\Gamma$ are
\begin{align}\label{eq:HGpi}
&(\tilde{H}_F)^{\Gamma(\pi)}_{11} = \frac{\Omega}{2} + \left[-\left(n_\Gamma-\frac{1}{2}\right)\Omega + 3\gamma\mathcal{J}_0(\alpha f(r))\left(1-\frac{a_0^2}{4}k^2\right)\right],\\
&(\tilde{H}_F)^{\Gamma(\pi)}_{12} = i(-1)^{n_\Gamma}\frac{3\gamma}{8}\mathcal{J}_{2n_\Gamma-1}(\alpha f(r))(k_x + i(-1)^{n_\Gamma\mod3}k_y)^2a^2_0e^{-i(2n_\Gamma-1)m\phi}, \\
&(\tilde{H}_F)^{\Gamma(\pi)}_{22} = \frac{\Omega}{2} - \left[-\left(n_\Gamma-\frac{1}{2}\right)\Omega + 3\gamma\mathcal{J}_0(\alpha f(r))\left(1-\frac{a_0^2}{4}k^2\right)\right],
\end{align}
and at $K$ are
\begin{align}\label{eq:HKpi}
&(\tilde{H}_F)^{K(\pi)}_{11} = \frac{\Omega}{2} + \frac{1}{\eta^2+1}\left[\left\{\left(n_K+\frac{1}{2}\right)\eta^2 + \left(n_K - \frac{1}{2}\right)\right\}\Omega + 6\gamma\eta\mathcal{J}_1(\alpha f(r))\left(1 - \frac{k^2a_0^2}{4}\right)\right], \\
&(\tilde{H}_F)^{K(\pi)}_{12} = (-1)^{(n_K+1)}\frac{3\gamma}{2(\eta^2+1)}\mathcal{J}_{2n_K-1}(\alpha f(r))(k_x - i(-1)^{n_K\mod3}k_y)a_0e^{2in_Km\phi}, \\
&(\tilde{H}_F)^{K(\pi)}_{22} = \frac{\Omega}{2} - \frac{1}{\eta^2+1}\left[\left\{\left(n_K+\frac{1}{2}\right)\eta^2 + \left(n_K - \frac{1}{2}\right)\right\}\Omega + 6\gamma\eta\mathcal{J}_1(\alpha f(r))\left(1 - \frac{k^2a_0^2}{4}\right)\right],
\end{align}
where $\eta(r)$ is given by Eq.~\eqref{eq:eta}. We have pseudo-OAM operators $\hat l^{\Gamma(\pi)}$ and $\hat l^{K(\pi)}$ given by
\begin{align}\label{eq:lpi}
    \hat l^{\Gamma(\pi)} & = -i\partial_\phi + \left[\frac{(2n_\Gamma-1)}{2}m - (-1)^{n_\Gamma\mod3}\right], \\
  \hat l^{K(\pi)} & = -i\partial_\phi - \left(n_Km - \frac{(-1)^{n_K\mod3}}{2}\right).
\end{align}

Correspondingly, the eigenvalues of the pseudo-OAM operator are
\begin{align}\label{eq:levalpi}
&l_\pm^{\Gamma(\pi)} = l \mp \left[\frac{(2n_\Gamma-1)}{2}m - (-1)^{n_\Gamma\mod 3}\right], \\
&l_\pm^{K(\pi)} = l \pm \left(n_Km - \frac{(-1)^{n_K\mod 3}}{2}\right).
\end{align}
for $\Gamma$ and $K$ respectively. Thus, from the effective Hamiltonians we can see that there is a quadratic band crossing at $\Gamma$ and linear band crossing at $K$ point. The complete Floquet eigenstates for $\Gamma$ and $K$ are
\begin{align}
\psi_{l}^{\Gamma(\pi)}(\mathbf{r},t) &= e^{il_+^{\Gamma(\pi)}\phi}w^{\Gamma(\pi)}_+(r) u^{(1-n_\Gamma)}_+(\Gamma,\mathbf{r},t) \; + \; e^{il_-^{\Gamma(\pi)}\phi}
w^{\Gamma(\pi)}_-(r) u^{(n_\Gamma)}_-(\Gamma,\mathbf{r},t), \label{eq:psipiG}\\
\psi_{l}^{K(\pi)}(\mathbf{r},t)&= e^{il_+^{K(\pi)}\phi}w^{K(\pi)}_+(r) u^{(n_K)}_+(K,\mathbf{r},t) \; + \; e^{il_-^{K(\pi)}\phi} w^{K(\pi)}_-(r) u^{(-n_K)}_-(K,\mathbf{r},t).\label{eq:psipiK}
\end{align}

\section{Analytical results for degenerate bound states}\label{app:wbound}
Consider a crossing between $n_\Gamma = \pm 1$ modes at the Floquet zone center. Using Eqs.~\eqref{eq:HG0-11} and~~\eqref{eq:HG0-12} to the lowest order in $\vex k$, we have
\begin{equation}\label{eq:HG01}
    \tilde{H}_F^{\Gamma(0)} = \mu_\Gamma(r)\sigma_z + \frac{3\gamma}{2}\mathcal{J}_2(\alpha f(r))a_0\left[-i(k_x - ik_y)e^{-2im\phi}\sigma_+ + \text{h.c.}\right], 
\end{equation}
where the dynamically generated mass $\mu_\Gamma(r) = -\Omega + 3\gamma\mathcal{J}_0(\alpha f(r))$ and $\sigma_+ = (\sigma_x + i\sigma_y)/2$ . The mass term vanishes at $r=r_c$ and changes sign on either side of the gap closing. The bound-state solutions are found by solving the Schr\"odinger equation $\tilde H_F^{\Gamma(0)} \begin{pmatrix}w_+^{\Gamma(0)} \\ w_-^{\Gamma(0)} \end{pmatrix} = 0$ after replacing $\mathbf{k}\to-i\bm\nabla_r$. The radial wavefunctions $w_\pm^{\Gamma(0)}(r)$ satisfy the differential equations,
\begin{equation} \label{eq:GZMC}
    \left[\partial_r \mp \frac{(l\mp 1/2)}{r} + \frac{1}{2\mathcal{J}_2(\alpha f(r))}\partial_r \mathcal{J}_2(\alpha f(r))\right]w_\pm^{\Gamma(0)} - \frac{2\mu_\Gamma(r)}{3\gamma\mathcal{J}_2(\alpha f(r))a_0}w_\mp^{\Gamma(0)} = 0.
\end{equation}

These equations are in general difficult to solve analytically, although we still can gain some information about the structure of $w_\pm^{\Gamma(0)}(r)$  by looking at the limits $r\rightarrow 0$ and $r\rightarrow r_c$ . As $r\rightarrow 0$, $\mu_\Gamma(r)\rightarrow$ constant and $\mathcal{J}_2(\alpha f(r)) \sim r^{2|m|}$  and Eq.~\eqref{eq:GZMC} becomes
\begin{equation}\label{eq:GZMCr0}
    r^{2|m|}\left[\partial_r \mp \frac{(l\mp 1)}{r}\right]w_\pm^{\Gamma(0)} - c_0w_\mp^{\Gamma(0)} = 0,
\end{equation}
where $c_0$ is a constant. Eq.~\eqref{eq:GZMCr0} can be equivalently written as
\begin{equation}\label{eq:GZMCr0appr}
    \left[\partial_r^2 + \frac{2(|m|+1)}{r}\partial_r - \frac{c_0^2}{r^{4|m|}}\right]w_\pm^{\Gamma(0)} = 0,
\end{equation}
where we have also used the approximation that $w_\pm^{\Gamma(0)}/r^2 \ll w_\pm^{\Gamma(0)}/r^{4|m|}$.  Eq.~\eqref{eq:GZMCr0appr} can be recast into the differential equation for the modified Bessel function after suitable substitutions, yielding
\begin{equation}\label{Gsol0}
    w_\pm^{\Gamma(0)}(r) \sim \frac{1}{r}\exp\left[-\frac{|c_0|}{(2|m|-1)r^{2|m|-1}}\right].
\end{equation}
From the form of the wavefunction we can see that as $r\rightarrow 0$, $w_\pm^{\Gamma(0)}(r)\rightarrow 0$ for all values of $m$ as expected. On the other hand, as $r\rightarrow r_c$, $\mu_\Gamma(r)\sim (r_c-r)$ and $\mathcal{J}_2(\alpha f(r))\rightarrow$ constant. 

We perform the substitution $w_\pm^{\Gamma(0)} = r^{\pm(l\mp 1/2)}v_\pm$ that transforms Eq.\eqref{eq:GZMC} into
\begin{equation}\label{eq:GZMrcapprox}
    \left[\partial_r^2 + \frac{1}{(r_c - r)}\partial_r - c_1^2(r_c-r)^2\right]v_\pm = 0 ,
\end{equation}
where $c_1$ is a constant and we have used the approximation  $|r-r_c| \ll r$ as $r\rightarrow r_c$ . Thus we find
\begin{equation}\label{eq:GZMsolrc}
    w_\pm^{\Gamma(0)}(r) \sim r^{\pm(l\mp 1/2)} [A - B(r_c-r)^2 + c_1^2(r_c-r)^4],
\end{equation}
where $A$ and $B$ are arbitrary constants. The above solution is peaked around $r=r_c$ and falls off on either side for all values of $l$ and $B>A$. 

Next we consider a crossing between the $n_K = 2,-3$ degenerate modes at the Floquet zone center. From Eqs.~\eqref{eq:HK0-11} and~\eqref{eq:HK0-12}, to the lowest order of $\vex k$, we have
\begin{equation}\label{eq:HK02}
    \tilde{H}_F^{K(0)} = \mu_K(r)\sigma_z + \frac{3}{2(\eta^2+1)}\gamma\mathcal{J}_4(\alpha f(r))a_0\left[-i(k_x+ik_y)e^{5im\phi}\sigma_+ + \text{ h.c.}\right],
\end{equation}
where
\begin{equation}
\mu_K(r) = \bigg(\frac{3\eta^2 + 2}{\eta^2+1}\bigg)\Omega + \bigg(\frac{6\eta}{\eta^2+1}\bigg)\gamma\mathcal{J}_1(\alpha f(r)),
\end{equation}
and $\eta(r)$ given by Eq. ~\eqref{eq:eta}. In this case also, the mass term vanishes at $r = r_c$ and changes sign on either side. The bound-state solution at zero quasienergy satisfies the following differential equation for the wavefunctions $w^{K(0)}_\pm(r)$,
\begin{equation}\label{eq:KZM}
    \left[\partial_r \pm \frac{(l\pm 1/2)}{r} + \frac{1}{2\mathcal{J}_4(\alpha f(r))}\partial_r \mathcal{J}_4(\alpha f(r))\right]w_\pm^{K(0)} - \frac{2(\eta^2+1)\mu_K(r)}{3\gamma\mathcal{J}_4(\alpha f(r))a_0}w_\mp^{K(0)} = 0.
\end{equation}
Using similar arguments as before,
\begin{equation}\label{eq:KZMsol}
w_\pm^{K(0)}(r) \sim 
\begin{cases}
    \frac{1}{r}\exp\left[-\frac{|c'_0|}{(4|m|-1)r^{4|m|-1}}\right], & r\rightarrow 0\\
    r^{\pm(l\mp 1/2)} [A' - B'(r_c-r)^2 + c'^2_1 (r_c-r)^4], & r\rightarrow r_c
\end{cases}
\end{equation}
where $c'_0$ , $c'_1$ , $A'$ and $B'$ are constants and $B'>A'$. The spatial distribution of $w_\pm^{K(0)}(r)$ is similar to that of $w_\pm^{\Gamma(0)}(r)$.

The values of the constants $c_0$, $c_1$, $c'_0$ and $c'_1$ are fixed by the system parameters $\Omega$, $\alpha$, $|m|$ and $r_c$.

\section{Selection rules for $\Gamma\rightarrow K$ transition}\label{app:selection}
\subsection{Floquet zone center}
For the $\Gamma\rightarrow K$ transition, we firstly need a scattering potential $V(\mathbf{r})$ that connects $\Gamma$ and $K$, since they are at different momenta. Then in order to have a non-zero overlap $|\langle\psi^{K(0)}_l|V(\mathbf{r})|\psi^{\Gamma(0)}_l\rangle|$, we get appropriate conditions on $n_\Gamma,n_K$ as well as OAM, $m$. Using Eqs. ~\eqref{eq:psiG0} and ~\eqref{psiK0}, 
\begin{align}\label{overlap0}
 \langle\psi^{K(0)}_l|V(\mathbf{r})|\psi^{\Gamma(0)}_l\rangle = \int_0^\infty rdr \int_0^{2\pi} d\phi \int_0^{2\pi/\Omega} &dt \left[
 w_+^{\Gamma(0)}w_+^{K(0)}u_A^*e^{i(l_+^{\Gamma(0)}-l_+^{K(0)}-m)\phi}e^{-i(n_\Gamma+n_K+1)\Omega t} \right. \nonumber \\
 &+ w_-^{\Gamma(0)}w_+^{K(0)}u_A^*e^{i(l_-^{\Gamma(0)}-l_+^{K(0)}-m)\phi}e^{i(n_\Gamma-n_K-1)\Omega t} \nonumber \\
 &+ w_+^{\Gamma(0)}w_-^{K(0)}u_B^*e^{i(l_+^{\Gamma(0)}-l_-^{K(0)}-m)\phi}e^{-i(n_\Gamma-n_K)\Omega t} \nonumber \\
 &+ w_-^{\Gamma(0)}w_-^{K(0)}u_B^*e^{i(l_-^{\Gamma(0)}-l_-^{K(0)}-m)\phi}e^{i(n_\Gamma+n_K)\Omega t} \nonumber \\
 &- w_+^{\Gamma(0)}w_+^{K(0)}u_B^*e^{i(l_+^{\Gamma(0)}-l_+^{K(0)})\phi}e^{-i(n_\Gamma+n_K)\Omega t} \nonumber \\
 &+  w_-^{\Gamma(0)}w_+^{K(0)}u_B^*e^{i(l_-^{\Gamma(0)}-l_+^{K(0)})\phi}e^{i(n_\Gamma-n_K)\Omega t} \nonumber \\
 &- w_+^{\Gamma(0)}w_-^{K(0)}u_A^*e^{i(l_+^{\Gamma(0)}-l_-^{K(0)})\phi}e^{-i(n_\Gamma-n_K-1)\Omega t} \nonumber \\
 &+ \left. w_-^{\Gamma(0)}w_-^{K(0)}u_A^*e^{i(l_-^{\Gamma(0)}-l_-^{K(0)})\phi}e^{i(n_\Gamma+n_K+1)\Omega t}
 \right].
\end{align}
Since $n_\Gamma,n_K>0$, we see that there are two possible cases for nonvanishing overlaps: $n_\Gamma=n_K = 2\: \mod 3$ or $n_\Gamma=n_K+1 = 1 \mod 3$. 
In the former case, we must have either $l_+^{\Gamma(0)} = l_-^{K(0)} + m$, which yields $m = 2$ and $\langle\psi^{K(0)}_l|V(\mathbf{r})|\psi^{\Gamma(0)}_l\rangle = \int_0^\infty w_+^{\Gamma(0)}w_-^{K(0)}u_B^* rdr$, or $l_-^{\Gamma(0)} = l_+^{K(0)}$, which yields $m = -2$ and $\langle\psi^{K(0)}_l|V(\mathbf{r})|\psi^{\Gamma(0)}_l\rangle = \int_0^\infty w_-^{\Gamma(0)}w_+^{K(0)}u_B^* rdr$. 
Similarly, in the latter case, we must have either $l_-^{\Gamma(0)} = l_+^{K(0)} + m$,  which yields $m=2$ and $\langle\psi^{K(0)}_l|V(\mathbf{r})|\psi^{\Gamma(0)}_l\rangle = \int_0^\infty w_-^{\Gamma(0)}w_+^{K(0)}u_A^* rd.$, or $l_+^{\Gamma(0)} = l_-^{K(0)}$, which yields $m=-2$ and $\langle\psi^{K(0)}_l|V(\mathbf{r})|\psi^{\Gamma(0)}_l\rangle = -\int_0^\infty w_+^{\Gamma(0)}w_-^{K(0)}u_A^* rdr$.

\subsection{Floquet zone boundary}
For $\Omega/2$ quasienergy, we use Eqs. ~\eqref{eq:psipiG} and ~\eqref{eq:psipiK} to get a non-zero overlap 
\begin{align}\label{overlappi}
\langle\psi^{K(\pi)}_l|V(\mathbf{r})|\psi^{\Gamma(\pi)}_l\rangle = \int_0^\infty rdr \int_0^{2\pi} d\phi \int_0^{2\pi/\Omega} &dt \left[ w_+^{\Gamma(\pi)}w_+^{K(\pi)}u_A^*e^{i(l_+^{\Gamma(\pi)}-l_+^{K(\pi)}-m)\phi}e^{-i(n_\Gamma+n_K)\Omega t} \right. \nonumber \\ 
& + w_-^{\Gamma(\pi)}w_+^{K(\pi)}u_A^*e^{i(l_-^{\Gamma(\pi)}-l_+^{K(\pi)}-m)\phi}e^{i(n_\Gamma-n_K-1)\Omega t} \nonumber \\ 
& + w_+^{\Gamma(\pi)}w_-^{K(\pi)}u_B^*e^{i(l_+^{\Gamma(\pi)}-l_-^{K(\pi)}-m)\phi}e^{-i(n_\Gamma-n_K)\Omega t}  \nonumber \\ 
& + w_-^{\Gamma(\pi)}w_-^{K(\pi)}u_B^*e^{i(l_-^{\Gamma(\pi)}-l_-^{K(\pi)}-m)\phi}e^{i(n_\Gamma+n_K-1)\Omega t}  \nonumber \\ 
& - w_+^{\Gamma(\pi)}w_+^{K(\pi)}u_B^*e^{i(l_+^{\Gamma(\pi)}-l_+^{K(\pi)})\phi}e^{-i(n_\Gamma+n_K-1)\Omega t}  \nonumber \\ 
& + w_-^{\Gamma(\pi)}w_+^{K(\pi)}u_B^*e^{i(l_-^{\Gamma(\pi)}-l_+^{K(\pi)})\phi}e^{i(n_\Gamma-n_K)\Omega t}  \nonumber \\ 
& - w_+^{\Gamma(\pi)}w_-^{K(\pi)}u_A^*e^{i(l_+^{\Gamma(\pi)}-l_-^{K(\pi)})\phi}e^{-i(n_\Gamma-n_K-1)\Omega t}  \nonumber \\ 
& + \left. w_-^{\Gamma(\pi)}w_-^{K(\pi)}u_A^*e^{i(l_-^{\Gamma(\pi)}-l_-^{K(\pi)})\phi}e^{i(n_\Gamma+n_K)\Omega t}\right].
\end{align}
Since $n_\Gamma,n_K>0$, we see that there are two possible cases for nonvanishing overlaps: $n_\Gamma=n_K = 1\mod 3$ or $n_\Gamma=n_K+1 = 0 \mod 3$ . In the former case we must have either  $l_+^{\Gamma(\pi)} = l_-^{K(\pi)} + m$, which yields $m=-1$ and $\langle\psi^{K(\pi)}_l|V(\mathbf{r})|\psi^{\Gamma(\pi)}_l\rangle = \int_0^\infty w_+^{\Gamma(\pi)}w_-^{K(\pi)}u_B^* rdr$   or $l_-^{\Gamma(\pi)} = l_+^{K(\pi)}$, which yields $m = 1$ and $\langle\psi^{K(\pi)}_l|V(\mathbf{r})|\psi^{\Gamma(\pi)}_l\rangle = \int_0^\infty w_-^{\Gamma(\pi)}w_+^{K(\pi)}u_B^* rdr$. 
Similarly, in the latter case, we must have either $l_-^{\Gamma(\pi)} = l_+^{K(\pi)} + m$, which yields $m = -1$ and $\langle\psi^{K(\pi)}_l|V(\mathbf{r})|\psi^{\Gamma(\pi)}_l\rangle = \int_0^\infty w_-^{\Gamma(\pi)}w_+^{K(\pi)}u_A^* rdr$ or, $l_+^{\Gamma(\pi)} = l_-^{K(\pi)}$, which yields $m = 1$ and $\langle\psi^{K(\pi)}_l|V(\mathbf{r})|\psi^{\Gamma(\pi)}_l\rangle = -\int_0^\infty w_+^{\Gamma(\pi)}w_-^{K(\pi)}u_A^* rdr$.

\section{$\Gamma\rightarrow K$ transition process}\label{app:trans}
Let, the transition amplitudes be $b = |\langle\psi^{K(\varepsilon)}_l|V(\mathbf{r})|\psi^{\Gamma(\varepsilon)}_l\rangle|$ and $b' = |\langle\psi^{K'(\varepsilon)}_l|V(\mathbf{r})|\psi^{\Gamma(\varepsilon)}_l\rangle|$ where, $\varepsilon = 0$ ($m=\pm 2$) or $\varepsilon = \Omega/2$ ($m=\pm 1$). For a given quasienergy, the transition amplitudes are in general different and depend on the sign of OAM. The reason is that the potential $V(\mathbf{r})$ connects the wavefunctions at opposite pairs of sublattices for $\Gamma\rightarrow K$ and $\Gamma\rightarrow K'$ for a particular sign of OAM. For example, consider multiple degeneracies at the Floquet zone boundary with $m=+1$ and $n_\Gamma = n_K$,
\begin{align}
    &b = \left|\int_0^\infty rdr \; w_-^{\Gamma(\pi)}(r)w_+^{K(\pi)}(r)u_B^*\right|\label{ampGK}, \\
    &b' = \left|\int_0^\infty rdr \; w_+^{\Gamma(\pi)}(r)w_-^{K'(\pi)}(r)u_A^*\right|\label{GK'}.
\end{align}
Thus, OAM of light can distinguish between the $K$ and $K'$ points due to its non-trivial helicity. Then in the basis $\{\ket{\psi^{K(\varepsilon)}_l},\ket{\psi^{\Gamma(\varepsilon)}_l},\ket{\psi^{K'(\varepsilon)}_l}\}$, $V(\mathbf{r})$ can be written as,
\begin{equation}\label{V}
V(\mathbf{r}) =  \begin{pmatrix} 0 & b & 0 \\ b & 0 & b' \\ 0 &  b' & 0 \end{pmatrix}.
\end{equation}
The eigenvalues are $E = 0, \pm\sqrt{b^2+b'^2}$ and the corresponding eigenvectors are, 
$$\frac{1}{\sqrt{b^2+b'^2}}\begin{pmatrix} b' \\ 0 \\ -b \end{pmatrix} \quad , \quad \frac{1}{\sqrt{2(b^2+b'^2)}} \begin{pmatrix} b \\ \pm \sqrt{b^2+b'^2} \\ b' \end{pmatrix}.$$ So in the presence of $V(\mathbf{r})$, if we prepare our initial state at $\Gamma$ as, $\ket{\Psi(t=0)} = (0 \quad 1 \quad 0)^T$, then after time $t$, 
\begin{equation}
\ket{\Psi^{\Gamma\rightarrow K(K')}(t)} = \frac{1}{\sqrt{b^2+b'^2}}\begin{pmatrix} i\sin(\sqrt{b^2+b'^2}t)b \\ \cos(\sqrt{b^2+b'^2}t)\sqrt{b^2+b'^2} \\ i\sin(\sqrt{b^2+b'^2}t)b' \end{pmatrix}. 
\end{equation}
 Thus, when $t = \frac{(2\nu+1)\pi}{2\sqrt{b^2+b'^2}}$ ($\nu\in\mathbb{Z}$), $\ket{\Psi}$ fully transitions into a superposition of $K$ and $K'$, and when $t = \frac{\nu\pi}{\sqrt{b^2+b'^2}}$, $\ket{\Psi}$ transitions back to $\Gamma$. Similarly, if we start in a superposition of $K$ and $K'$, $\ket{\Psi(t=0)} = (b \quad 0 \quad b')^T/\sqrt{b^2+b'^2}$, then after time $t$, 
 \begin{equation}
 \ket{\Psi^{K(K')\rightarrow\Gamma}(t)} = \frac{1}{\sqrt{b^2+b'^2}}\begin{pmatrix} \cos(\sqrt{b^2+b'^2}t)b \\ i\sin(\sqrt{b^2+b'^2}t)\sqrt{b^2+b'^2} \\ \cos(\sqrt{b^2+b'^2}t)b' \end{pmatrix}.
\end{equation} 
Thus, when $t = \frac{(2\nu+1)\pi}{2\sqrt{b^2+b'^2}}$ ($\nu\in\mathbb{Z}$), $\ket{\Psi}$ fully transitions into $\Gamma$, and when $t = \frac{\nu\pi}{\sqrt{b^2+b'^2}}$, $\ket{\Psi}$ transitions back to a superposition of $K$ and $K'$. So even though we have used a three state basis, it is effectively a two-level system since $K$ and $K'$ points are at time-reversal invariant momenta. Therefore, using non-zero OAM and a suitable scattering potential $V(\mathbf{r})$ we have engineered a method to create and manipulate superpositions of the multiple degenerate bound states at the $\Gamma$ and $K(K')$ points.

\twocolumngrid

\bibliography{ref-v2}

\end{document}